\documentclass{mem}
\usepackage{natbib}\usepackage{txfonts}\usepackage{balance}
\usepackage{graphicx}
\usepackage[a4paper,breaklinks,dvipdfm]{hyperref}
\idline{86}{100}
\begin{document}
\def\teff{$T\rm_{eff }$}
\def\kms{$\mathrm {km s}^{-1}$}

\title{Solar diameter, eclipses and transits:
the importance of ground-based observations
}

   \subtitle{}

\author{
Costantino \,Sigismondi\inst{1} 
}

  \offprints{Costantino Sigismondi}

\institute{ICRA, Ateneo Pontificio Regina Apostolorum and Sapienza University of Rome, Italy;   visiting at Observatorio Nacional, Rio de Janeiro, Brazil. 
\email{sigismondi@icra.it}
}

\authorrunning{Sigismondi}

\titlerunning{Solar diameter from eclipses and transits}

\abstract{The lifetime of a satellite is limited, and its calibration may be not stable, 
it is necessary to continue ground-based measures of the solar diameter with
methods less affected by atmospheric turbulence, and optical aberrations:
planetary transits and total eclipses.
The state of art, advantages and limits of these methods are considered.

}
\maketitle{}

\section{Introduction: the solar variability}

The variability of the Sun is associated with its 11-years activity, but
it has been shown 
that these cycles are modulated over the centuries,
with variations in the solar luminosity.
Milankovitch shown other climate forcings acting over millenia, related to the Earth's orbit characteristics \citep{Usoskin}. 
Nevertheless while the irradiance variation is accurately verificated over the last three 11-years cycles, thanks to the radiometers onboard the ACRIM satellites, no sure evidence is available before 1980.
Therefore the solar activity modulations over several cycles are not quantitatively connected with the solar luminosity.
One way to know this connection is to measure the variations of the solar diameter.
The Stefan-Boltzmann law, with no temperature variations yields the relation $\Delta L/L= 2 \Delta R/R$.
Monitoring the variations of the solar radius, and cross-correlating them with the variations of the total solar irradiance we obtain the coefficient W=${\Delta R/R}\over{\Delta L/L}$, to be used to recover past values of the luminosity, through the radius from ancient eclipses and transits \citep{Sofia}.

\begin{figure}[t!]
\resizebox{\hsize}{!}{\includegraphics[clip=true]{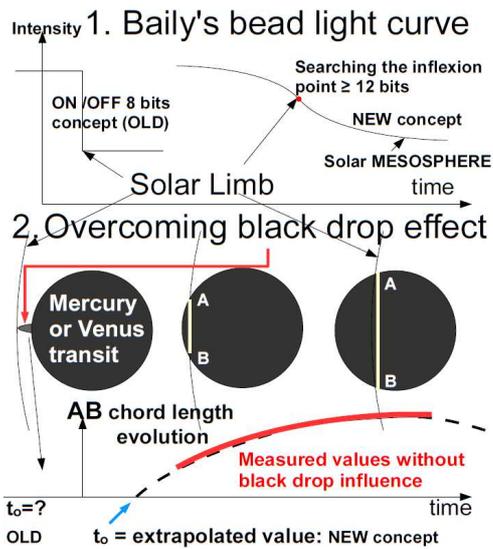}}
\caption{
\footnotesize
Old and new concepts in eclipses and planetary transits data analysis.
Baily's beads should be recorded with a photometric resolution of 12 bits, 
an acquisition rate larger than 100 frames per second, and eventually the possibility 
to record two or three color channels, in addition to the white light 
(the eclipse can be projected on a screen and
observed without neutral density filters). All these devices are already on the market.
The scan of the Limb Darkening Function LDF near the solar limb is made by the progress of the lunar limb 
at the angular speed of 0.3-0.5 arcsec/s. With a few beads observed with these new standards, some millisecond of accuracy in the solar diameter's determination are obtained.}
\label{sketch}
\end{figure}

\section{Solar diameter from eclipse data}

The observations of Baily's beads 
during central eclipses has been exploited to measure 
the solar diameter. The larger is the number of beads the more accurate is the 
determination of the solar diameter. To get more beads they are observed from the limits of the shadow band.
The profile of the lunar limb, the solar standard radius $R_{\odot}$=959.63 arsec and the ephemerides allow to predict the appearance of the beads.
All departures from these predictions are averaged to obtain the correction to $R_{\odot}$ \citep{Sigismondi09a}.
Kaguya, a lunar japanese probe, completed in 2009 the lunar limb survey with accuracy of $\pm 1$m, greatly improving our knowledge with respect to the photographic atlas of C. B. Watts published in 1962 with a 0.2 arscec (400 m on the Moon) of random error.  
The implement and test of this new dataset in the analyses of solar eclipses and lunar occultations are still ongoing. 

Solar eclipses occur each one or two years. Nowadays they are observed mainly by amateurs, with portable equipments settled in a few days or hours thanks to GPS alignement methods. In the past centuries the astronomers required several months to travel, to settle their instruments, and to test them.
IOTA, International Occultation Timing Association, is the organization of amateurs and professionals dedicated to this kind of observations. The discussions at the Four Centennial Clavius Meeting and XXXI ESOP of IOTA European Section held in Pescara on 24-27 August 2012, {\rm www.icranet.org/clavius2012} contributed also to the development of this topic \citep{JOA}. 
\\
For an eclipse, the observation is seldom longer than 5 minutes, even at the shadow's edges.
Therefore it is possible to complete the record of the Baily's beads even with a not perfect equatorial alignment of the telescope, or with a smooth manual tracking. For planetary transits, lasting a few hours, the alignment is more demanding.
\\
{\bf The Baily's beads technique} is an improvement with respect to the simple measurement of the duration of totality: in effect the totality is determined by the disappearance of the last bead at second contact and the reappearance of the first bead at third contact.
In the first 4 decades this technique consisted into the identification of the ON/OFF timing of the appearance or disappearance of the beads. These times were associated to the bead's location on the lunar limb, through its axis angle. The Atlas of Baily's beads gathers all eclipse data from 2005 to 2008 \citep{Sigismondi09b}.
\\
Thanks to this effort we could verify soon that observations made with different equipments yield different corrections to $R_{\odot}$.
This discrepancy is originated by the presence in the solar mesosphere of many emitting lines, whose blend is white light, located $\sim$0.7 arcsec above the photosphere. That white light was perceived as the continuation of the photospheric bead, of which we had to measure the ON/OFF timing. The problem was to understand when the bead's photospheric light ends.
The solution of this controversy is possible in two ways: spectroscopically by identifying the emission lines with respect to the continuum \citep{Bazin},  
and photometrically by identifying the inflexion point of the limb darkening function \citep{Raponi}.
Both the two solutions are object of research studies in collaboration with French CNRS and IOTA. 
The simplicity of the photometric method leave it suitable for amateurs observing missions, while the spectroscopic one is more complicated.
But the requirements of the video are nevertheless rather demanding.
Because of the need to identify the inflexion point, which is the standard definition of the solar limb \citep{Hill},
it is necessary to have unsaturated light curves of the Baily's beads from its appearance to its merging with the surrounding beads.
8-bits dynamics of standard CCD cameras, as the B/W Watec models, if not correctly filtered, saturates too early to see the inflexion point, while 16-bits cameras produce a large amount of data difficult to be recorded with on-field equipments.
\\
Similarly the standard acquisition rate is still limited to 25 to 30 frames per second, while camcorders with 60, 300 fps or even 1200 fps are already on the market.
Increasing the acquisition rate corresponds to an enlargement of data size, but in this way the video resolution becomes competitive with photometers, with the great advantage to have the spatial information of the incoming photons.
\\
CNRS eclipse missions to Hao Island in Polinesia (2010) and to Australia (2012), were equipped with an array of photometers displaced perpendicularly to the centerline, in order to monitor the duration of totality in different places, and to recover the solar diameter by using very accurate timings and the data are now under analysis.
\\
The historical data available on solar eclipses, before video era, are ON/OFF timings, and the analysis of Baily's beads Atlas 2005-2008 will allow to evaluate the errorbar on $R_{\odot}$ of the ancient visual eclipses data, useful through the W parameter for knowing the past solar irradiance. They are crucial for understanding the behaviour of solar diameter during the last 5 centuries, starting from the first annular eclipse recorded in the modern times, the one of Clavius in 1567, observed in Rome.
Moreover the utility of these observations is of great importance in order to bridge data from different satellites which still need absolute calibration methods.

\section{Solar diameter from Venus and Mercury transit data}

An absolute method of pixel calibration for satellite images should be independent 
on the focal length and the optical aberrations of the instrument.
From ground the atmospheric seeing can produce irregular oscillations in the light curve of a Baily's bead,
and it may affect the eclipses timing, especially when the signal is faint, but the alignment of Moon Sun and the observer is determined only the orbital motions. The same considerations apply with planetary transits.
\\ 
Regarding {\bf the black drop phenomenon}, commonly associated with planetary transits,
it applies also to the Baily's beads.
It is due to the interplay between the limb darkening function LDF, rapidly dropping to zero near the limb,
and the point spread function PSF of the telescope, in presence of a sharp cutting edge as the lunar edge or the planetary disk cast over the solar photosphere \citep{pasa}.
Because of the PSF-LDF convolution, especially if the observed point is off-axis and the image is out of focus, a black drop connects the two dark sides: the Moon or the planet and the black sky around the Sun \citep{horn}.
Especially when dealing with beads' merging, the timing is strongly affected by a {\sl black drop}-type effect.

The method of measuring the solar diameter with planetary transits consists to measure as precisely as possible the contact timings.
With the ephemerides we compute the timings for a standard solar radius ($R_{\odot}$=959.63 arcsec at 1 AU, i.e. exactly the angle formed by 696000 Km seen from 1 AU, adopted by the International Astronomical Union).
The ratio between the duration of the observed transit and the calculated one gives the corresponding correction to this standard radius.

To overcome the {\sl black drop} effect in the planetary transits 
is possible by measuring the lengths of the chords cut by the disk of the planet over the limb of the Sun and
to plot their evolution along time \citep{tesi}.
The analytical function describing this chord, without aberrations and {\sl black drop}, is used to extrapolate the time when the chords are zero, i.e. the contact timings, using the values measured far from the critical phases when the {\sl black drop} starts to occur. 
The {\sl black drop} is overcome also by using perfect optical systems in the best focus condition, as in the satellite images from HINODE or TRACE \citep{Wang}, the same occurred with the R. Dunn telescope, the Dutch Open Telescope and THEMIS.
Small telescopes of diameter around 6 cm give more problems, because of a strong diffractive effect, i.e. the main source for {\sl black drop} phenomenon.
The frequency of the transits of Mercury is about 13 per century, and they are really usefull for long term studies of the solar diameter. The last two transits of Venus have been observed in order to measure the solar diameter overcoming the {\sl black drop} effect.
In 2004 A. Ayomamitis took 50 images chronodated, 25 for the first contact and 25 for the last one, taken each minute with a 6 cm telescope with a filter centered in the $H_{\alpha}$ line at 256 levels of intensity (8-bits). 
In 2012 we organized an observational champaign at Huairou Solar Observing Station near Beijing with 1 image per second at a 28 cm telescope, with 4096 levels of intensity or 12-bits dynamics \citep{Xie}.
The data are still under examination, in collaboration with the VENUSTEX project \citep{tanga}. 
dedicated to the study of the aureole of Venus.
The refracted light through the upper Venus atmosphere tends to reduce the observed diameter of the planet and affects the deduced solar diameter by delaying the third contact and anticipating the second one. Therefore the internal contact timings are more separated because of the aureole, and the corresponding solar diameter is perceived as larger of the same angular amount of the aureole.
The possibility to study the aureole with special coronographs, when it is casted above the inner corona, which appears as black sky in normal eclipses video of Baily's beads, gives to us the idea of the effect of that aureole at the internal and external contact times.
For the transits of Mercury the problem of the atmosphere does not occur: but the planet is 13 arcsec of diameter at maximum, instead of 60 arcsec of Venus, and always under the influence of {\sl black drop}.  

\section{Conclusions}
The lifetime of satellites is very short with respect to the timescale of solar variations.
Instrumental aberrations varying with time modify the results of satellites, and they can not be repeteadly calibrated. 
Ageing of detectors onboard satellites is the other critical factor for these measurements, limitating the duration of the analysis. 
Moreover the decay of the orbit around the Earth of satellites reduces their operativity.
On the lagrangian point L2, as for the case of Solar Heliospheric Observer (SOHO), there is no decay.
Meanwhile the Solar Dynamics Observer (SDO) was placed in geosynchronous orbit to enable an extremely high downlink data rate, not available at L2.
Conversely the timing accuracy of the ground-based observations of total eclipses and planetary transits here considered can attain very high accuracies.
Before the studies presented in this paper the method of the eclipses was limited by the concepts of step function implicitly associated with LDF.
The method of planetary transit was heavily limited by the {\sl black drop} effect.
In the last transits of Venus we studied the possibility to overcome this problem.
A preliminary result on 2004 transit shown the influence of atmospheric seeing on the contact timings of the transit.
More accuracy are expected by the analysis of the 2012 data, of much better quality.
On 2016 the next transit of Mercury visible from Europe will be the next occasion to measure the solar diameter at milliarcescond level from ground with such methods based on timing.

\bibliographystyle{aa}

\begin{thebibliography}{}

\bibitem[Usoskin(2008)]{Usoskin}
Usoskin, I., \ 2008 Living Rev. Sol. Phys. 5, 3

\bibitem[Sofia \& Chan (1982)]{Sofia} 
Sofia, S. \& Chan K., \ 1982 Sol.Phys. 76, 145 

\bibitem[Beisker (2012)]{JOA} 
Beisker, W. \ 2012 J.Occ.Astron., 4, 14-19 

\bibitem[Sigismondi(2009a)]{Sigismondi09a}
Sigismondi, C., \ 2009 Sci.China G 52, 1773

\bibitem[Sigismondi et al.(2009b)]{Sigismondi09b}
Sigismondi, C., et al., \ 2009 Sol.Phys. 258, 191

\bibitem[Sigismondi et al.(2011)]{Bazin}
Sigismondi, C., et al.,\ 2011 arXiv 1106.2197

\bibitem[Raponi et al.(2012)]{Raponi}
Raponi, A., et al., \ 2012 Sol.Phys. 278, 269

\bibitem[Hill et al.(1975)]{Hill}
Hill, H. A., et al., \ 1975 \apj 200, 484

\bibitem[Pasachoff et al.(2011)]{pasa}
Pasachoff, J. M., et al., \ 2011 \aj 141, 112

\bibitem[Horn d'Arturo(1922)]{horn} 
Horn d'Arturo, G., \ 1922 Pub. Oss. Astron. R. Univ. Bologna, vol. I, n.3

\bibitem[Sigismondi(2013)]{tesi}
Sigismondi, C.,\ 2013 arXiv 1301.0311 

\bibitem[Wang \& Sigismondi(2012)]{Wang}
Wang, X. \& Sigismondi, C., \ 2012 
arXiv 1210.8286 

\bibitem[Xie et al.(2012)]{Xie}
Xie, W., et al., \ 2012 
arXiv 1210.8451 

\bibitem[Tanga et al.(2012)]{tanga} 
Tanga, P., et al., \ 2012 ICARUS, 218, 207

\end{thebibliography}

\end{document}